\documentclass[aps,prb,twocolumn,groupedaddress,showpacs,amsmath,amssymb,amsfonts,floatfix]{revtex4}

\usepackage{graphicx,bm}

\begin{document}
\title{Scaling Properties of Fidelity in Spin-one Anisotropic Model}

\author{Yu-Chin Tzeng}
\affiliation{Department of Physics, Tunghai University, Taichung,
Taiwan}

\author{Min-Fong Yang}
\email{mfyang@thu.edu.tw} %
\affiliation{Department of Physics, Tunghai University, Taichung,
Taiwan}

\date{\today}

%%%%%%%%%%%%%%%%%%%%%%%%%%%%%%%%%%%%%%%%%%%%%%%%%%%%%%%%%%%%%%%%%%%
\begin{abstract}
By means of the density matrix renormalization group technique,
the scaling relation of the fidelity susceptibility proposed
recently is verified for the spin-one $XXZ$ spin chain with an
on-site anisotropic term. Moreover, from the results of both the
fidelity susceptibility and the entanglement entropy, the critical
points and some of the corresponding critical exponents are
determined through a proper finite-size scaling analysis, and
these values agree with the findings in the literature. Thus our
work provides a numerical support of the use of the fidelity in
detecting quantum phase transitions.
\end{abstract}
%%%%%%%%%%%%%%%%%%%%%%%%%%%%%%%%%%%%%%%%%%%%%%%%%%%%%%%%%%%%%%%%%%%

%68.35.Rh,          %Phase transitions and critical phenomena
%05.70.Jk,          %Critical point phenomena
%64.60.-i           %General studies of phase transitions
%75.30.kz,          %magnetic phase boundaries (including magnetic transitions, metamagnetism, etc)
%05.30.-d           %Quantum statistical mechanics

\maketitle

%%%%%%%%%%%%%%%%%%%%%%%%%%%%%%%%%%%%%%%%%%%%%%%%%%%%%%
%\section{Introduction}
%%%%%%%%%%%%%%%%%%%%%%%%%%%%%%%%%%%%%%%%%%%%%%%%%%%%%%

Quantum phase transitions (QPTs),~\cite{Sachdev:book} driven by
purely quantum fluctuations, are characterized by the dramatic
changes in the ground state of a many-body system as the
controlling parameters in the system Hamiltonian are varied across
critical points. Due to latest advances in quantum information
science,~\cite{Nielsen:book} people attempt to characterize QPTs
from the perspective of quantum information. One of the
well-studied aspects is to explore the role of quantum
entanglement in identifying QPTs.~\cite{Amico0703044} In
particular, as a bipartite entanglement measure, the entanglement
entropy of a block of length $l$ for one-dimensional systems is
shown to exhibit qualitatively different scaling behaviors at and
off
criticality.~\cite{Holzhey,Vidal03,Korepin04,CC04,FKR,LSCA,ZBFS,Fan07,Katsura07,Cardy0708.2978}
The entanglement entropy saturates to a finite bound as the length
$l$ increases for noncritical (gapped)
systems,~\cite{Vidal03,FKR,Katsura07} whose value can vary for
different boundary conditions.~\cite{Fan07} However, the
entanglement entropy increases logarithmically for critical
(gapless)
systems.~\cite{Holzhey,Vidal03,Korepin04,CC04,LSCA,ZBFS,Cardy0708.2978}
By using conformal field theory, a universal scaling is expected
at a quantum critical point, and its expression depends again on
the boundary conditions. Thus the divergent character of the
entanglement entropy in the finite-size scaling can faithfully
indicate the existence of the critical points for one-dimensional
systems.

In the last few years, the ground-state
fidelity~\cite{HTQuan2006,Zanardi06} (and its second derivatives,
the so-called ``fidelity susceptibility"~\cite{YLG07}), another
concept emerged from quantum information science, attracts much
attention on their application to the analysis of
QPTs.~\cite{Zanardi06,YLG07,ZCG0606130,Cozzini07,CIZ0611727,Buonsante07,Oelkers07,ZGC0701061,CVZ07,CWGW0706.0072,GKNL0706.2495,zhou,Yang07}
As illustrated before in several concrete models, it seems that
the singularity in the fidelity susceptibility can be an effective
tool in detecting critical points. Quite recently general scaling
analyses of the fidelity susceptibility are
proposed.~\cite{CVZ07,GKNL0706.2495} As explicitly shown in
Ref.~\onlinecite{CVZ07}, the fidelity susceptibility ${\cal S}$
must be bounded above in the thermodynamical limit for noncritical
(gapped) systems containing only local operators. However, for
critical (gapless) systems of finite size $L$, it fulfills scaling
relations
\begin{equation}
{\cal S}\sim L^{-\Delta_{Q}} \; , \qquad
\Delta_{Q}=2\Delta_{V}-2z-d \; ,  \label{scaling_S}
\end{equation}
where $d$ is the spatial dimension, $z$ is the dynamic exponent,
and $\Delta_{V}$ is the scaling dimension of the
transition-driving term in the Hamiltonian. This result implies
that the QPTs at those critical points with $\Delta_{Q}<0$ can be
detected by the power-law divergent behaviors in ${\cal S}$.

In this paper, the spin-one $XXZ$ spin chain with a uniaxial
single-ion anisotropic term is investigated, and we focus our
attention on the verification of the predicted scaling behavior of
the fidelity susceptibility in Eq.~(\ref{scaling_S}). It is known
that, while the QPTs can in principle be unveiled by the knowledge
of the entanglement entropy and the ground state fidelity, they
are usually difficult to be calculated due to the lack of
knowledge of the exact ground state wavefunctions. Although
numerical exact diagonalization can always be employed to evaluate
the entanglement entropy and the fidelity for small systems, this
method may not be able to demonstrate the scaling behaviors
because of finite-size effects.  Thus we need to resort to the
density matrix renormalization group (DMRG) technique~\cite{DMRG}
for the calculations for systems of large sizes. In the present
work, both the entanglement entropy and the fidelity
susceptibility are evaluated by means of the finite-system DMRG
technique under open boundary conditions for system sizes up to
$L=160$. In our DMRG calculations, up to 300 states per block are
kept and five DMRG sweeps are performed for the truncation error
being about $10^{-10}$. We find that developing peaks do appear in
both measurements, which signal precursors of the QPTs. Applying a
proper finite-size scaling analysis, the proposed scaling relation
in Eq.~(\ref{scaling_S}) is confirmed numerically. Besides, the
critical points in the thermodynamic limit and some of the
corresponding critical exponents are determined through a proper
finite-size scaling analysis, and these values agree with the
results in the literature. Moreover, the results coming from both
the entanglement entropy and the fidelity susceptibility are
consistent each other. This implies that both measurements are
equally suited for revealing QPTs and pinning down the critical
points in the present case.

%%%%%%%%%%%%%%%%%%%%%%%%%%%%%%%%%%%%%%%%%%%%%%%%%%%%%%
%\section{model Hamiltonian}
%%%%%%%%%%%%%%%%%%%%%%%%%%%%%%%%%%%%%%%%%%%%%%%%%%%%%%

The Hamiltonian for spin-one $XXZ$ spin chains of $L$ sites with
an on-site anisotropic term is
\begin{equation}
H=\sum_{j=1}^{L-1}\left( S_{j}^{x}S_{j+1}^{x} +
S_{j}^{y}S_{j+1}^{y} + \lambda S_{j}^{z}S_{j+1}^{z}\right) + D
\sum_{j=1}^{L}\left(S_{j}^{z}\right)^{2} ,\label{hamilt}
\end{equation}
where $S_j^\alpha$ ($\alpha=x,y,z$) are the spin-one operators at
the $j$-th lattice site. $\lambda$ and $D$ parametrize the
Ising-like and the uniaxial single-ion anisotropies, respectively.
The full phase diagram consists of six different
phases~\cite{denNijs89,Schulz86} (see
Refs.~\onlinecite{Chen03,DegliEspostiBoschi03,CamposVenuti06-2}
for recent numerical determinations). On the $\lambda>0$
half-plane, there consists of three phases known as the Haldane,
the large-$D$, and the N\'{e}el phases. All these three phases
show a nonzero energy gap above the ground state. Between these
phases, various types of phase transitions take place. There are a
Gaussian transition between the Haldane and the large-$D$ phases,
and an Ising transition between the N\'{e}el and the Haldane
phases. These two transitions merge at a tricritical point
$\lambda\simeq 3.20$ and $D\simeq
2.90$,~\cite{Chen03,DegliEspostiBoschi03} where the Haldane phase
disappears and the N\'{e}el-large-$D$ transition becomes first
order. Here we consider only the Gaussian and the Ising
transitions at $\lambda=1$. In this case, it has been found that,
as $D$ is decreased from a large value, one first meets a Gaussian
transition from the large-$D$ to the Haldane phases at the
critical point $D_c\simeq 0.99$, and then an Ising transition from
the Haldane to the N\'{e}el phases at $D_c\simeq
-0.31$.~\cite{Chen03,CamposVenuti06-2}

For the convenience of the following discussions, some details of
the Ising and the Gaussian transitions are
reviewed.~\cite{Sachdev:book} Both transition lines are of second
order with a dynamic exponent $z=1$. Nevertheless, the former is
described by a conformal field theory (CFT) with a central charge
$c=1/2$, while the latter by a $c=1$
CFT.~\cite{DegliEspostiBoschi03} Moreover, their singular
behaviors with the universality class of the transition, i.e.,
critical exponents, can be different. For the Ising transition, it
is known that the correlation length critical exponent $\nu=1$ and
the scaling dimension of the transition-driving term in the
Hamiltonian $\Delta_{V}=1$.~\cite{Sachdev:book} However, for the
Gaussian transition between the Haldane and the large-$D$ phase,
it is found that the low-energy effective continuum theory can be
described by the sine-Gordon
model~\cite{DegliEspostiBoschi03,note1}
\begin{equation}
H_{SG}=
\frac{1}{2}\left[\Pi^{2}+\left(\partial_{x}\Phi\right)^{2}\right]
-\frac{\mu}{a^{2}}\cos\left(\sqrt{4\pi K}\Phi\right) \; ,
\label{eq:SG}
\end{equation}
where $\Pi$ and $\Phi$ are the conjugate bosonic phase fields, and
$a$ is a short-distance cut-off of the order of the lattice
spacing. The coefficient $\mu\propto(D-D_c)$ in the vicinity of
the critical point $D_c$ for a given $\lambda$, and thus becomes
zero at the transition point. The value of the Luttinger liquid
parameter $K$ varies continuously between 1/2 and 2 along the
critical line. We note that all the scaling dimensions and the
critical exponents are determined by a single parameter $K$.
Consequently, they change continuously along the critical line.
From the sine-Gordon theory,~\cite{lukyanov97} it is found that
the critical exponent of the correlation length $\nu=1/(2-K)$ and
the scaling dimension $\Delta_{V}=K$ for the transition-driving
term $\cos(\sqrt{4\pi K}\Phi)$.

%%%%%%%%%%%%%%%%%%%%%%%%%%%%%%%%%%%%%%%%%%%%%%%%%%%%%%
%\section{DMRG results}
%%%%%%%%%%%%%%%%%%%%%%%%%%%%%%%%%%%%%%%%%%%%%%%%%%%%%%

%%%%%%%%%%%%%%%%%%%%%%%%%%%%%%%%%%%%%%%%%%%%%%%%%
\begin{figure}
\includegraphics[width=3in]{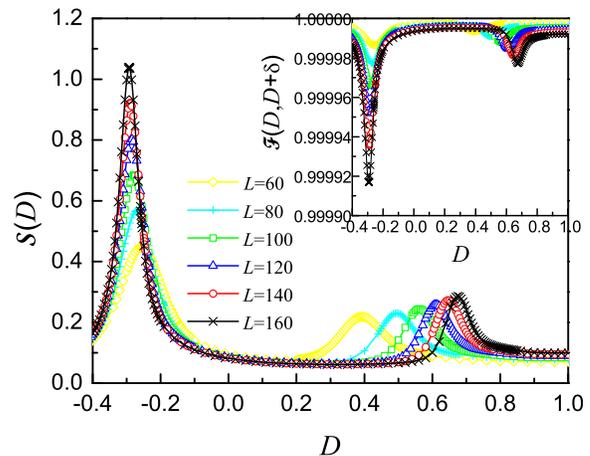}
\caption{(Color online) Fidelity susceptibility ${\cal S}(D)$ for
the spin-1 $XXZ$ spin chain in Eq.~(\ref{hamilt}) as functions of
$D$ for various sizes $L$ with $\lambda=1$. Inset shows the
fidelity ${\cal F}(D, D+\delta)$ as functions of $D$ for the
corresponding sizes. Here we take $\delta=10^{-3}$.} \label{fig:S}
\end{figure}
%%%%%%%%%%%%%%%%%%%%%%%%%%%%%%%%%%%%%%%%%%%%%%%%%

In the following, our DMRG results are presented in
order.~\cite{note2} The findings of the fidelity susceptibility
${\cal S}(D)$ and the ground state fidelity ${\cal F}(D,
D+\delta)$ are shown in Fig.~\ref{fig:S}. The fidelity
susceptibility, or the second derivative of the fidelity, is
calculated by~\cite{Cozzini07,Buonsante07}
\begin{equation}
{\cal S}(D)= \lim_{\delta \to 0} \frac{2 [ 1 - {\cal
F}(D,D+\delta)]}{L\;\delta^2} \; , \label{eq:fide_2_deri}
\end{equation}
where the ground-state fidelity (or the modulus of the overlap) is
given by~\cite{Zanardi06,note3}
\begin{equation}
{\cal F}(D, D+\delta) = |\langle\Psi_0(D)|\Psi_0(D+\delta)\rangle|
\label{eq:fide_def}
\end{equation}
with $|\Psi_0 (D)\rangle$ and $|\Psi_0 (D+\delta)\rangle$ being
two normalized ground states corresponding to neighboring
Hamiltonian parameters. In our calculations, $\delta=10^{-3}$ is
used. As shown in the inset of Fig.~\ref{fig:S}, drops in the
ground state fidelity are observed, which signal precursors of the
Gaussian and the Ising transitions in the model under
consideration. The drops in ${\cal F}(D, D+\delta)$ at the
right-hand side show the Gaussian transition, while those at the
left-hand side give the Ising one. Further evidences for
indicating QPTs are provided by the results of ${\cal S}(D)$. As
seen from Fig.~\ref{fig:S}, the maximum values ${\cal S}_{\rm
max}$ in the fidelity susceptibility grow with increasing size,
and thus indicate divergence in the $L\to\infty$ limit (see also
Fig.~\ref{fig:Smax} below). From the scaling analysis in
Ref.~\onlinecite{CVZ07}, these divergent behaviors in ${\cal S}$
must imply the appearance of the QPTs. Applying the finite-size
scaling, the critical points $D_c$ in the thermodynamic limit can
be determined from the locations $D_{\rm max}(L)$ of the local
maxima in ${\cal S}(D)$ on a size-$L$ system (see
Fig.~\ref{fig:crit_point} below).

\begin{figure}
\includegraphics[width=3in]{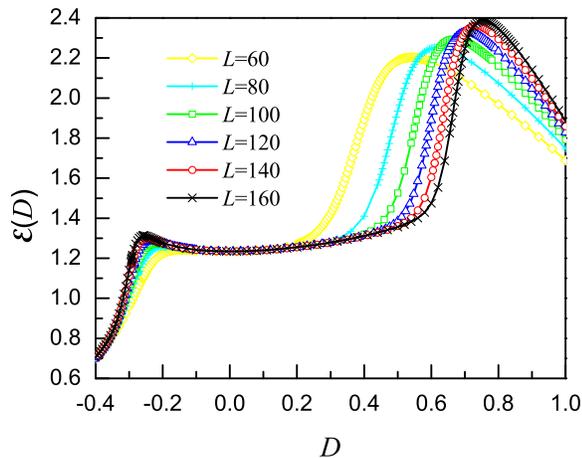}
\caption{(Color online) Entanglement entropy ${\cal E}(D)$ for the
spin-1 $XXZ$ spin chain in Eq.~(\ref{hamilt}) as functions of $D$
for various sizes $L$ with $\lambda=1$. } \label{fig:EE}
\end{figure}

As mentioned before, the divergent character of the entanglement
entropy can also show the existence of the QPTs. Thus the
entanglement entropy is evaluated for comparison. Here we consider
the entanglement entropy, or the von Neumann entropy of the
reduced density matrix $\rho_R(D)$ of the right-hand block of
$L/2$ contiguous spins
\begin{equation}
{\cal E}(D) = -{\rm Tr} \left[ \rho_R(D) {\rm log}_2 \rho_R(D)
\right] \; . \label{eq:entropy}
\end{equation}
Our DMRG results are shown in Fig.~\ref{fig:EE}. It is found that,
far away from the critical points, ${\cal E}(D)$ has no size
dependence, as expected for the gapped phases. Nevertheless, two
peaks develop as size $L$ increases. Again, these peaks indicate
the existence of the Gaussian and the Ising transitions, and the
corresponding critical points $D_c$ can be deduced from the
locations $D_{\rm max}(L)$ of the local maxima in ${\cal E}(D)$ on
a size-$L$ system, as discussed below.

%%%%%%%%%%%%%%%%%%%%%%%%%%%%%%%%%%%%%%%%%%%%%%%%%
%%%%%%%%%%%%%%%%%%%%%%%%%%%%%%%%%%%%%%%%%%%%%%%%%
%%%%%%%%%%%%%%%%%%%%%%%%%%%%%%%%%%%%%%%%%%%%%%%%%
\begin{figure}
\includegraphics[width=3in,height=2in]{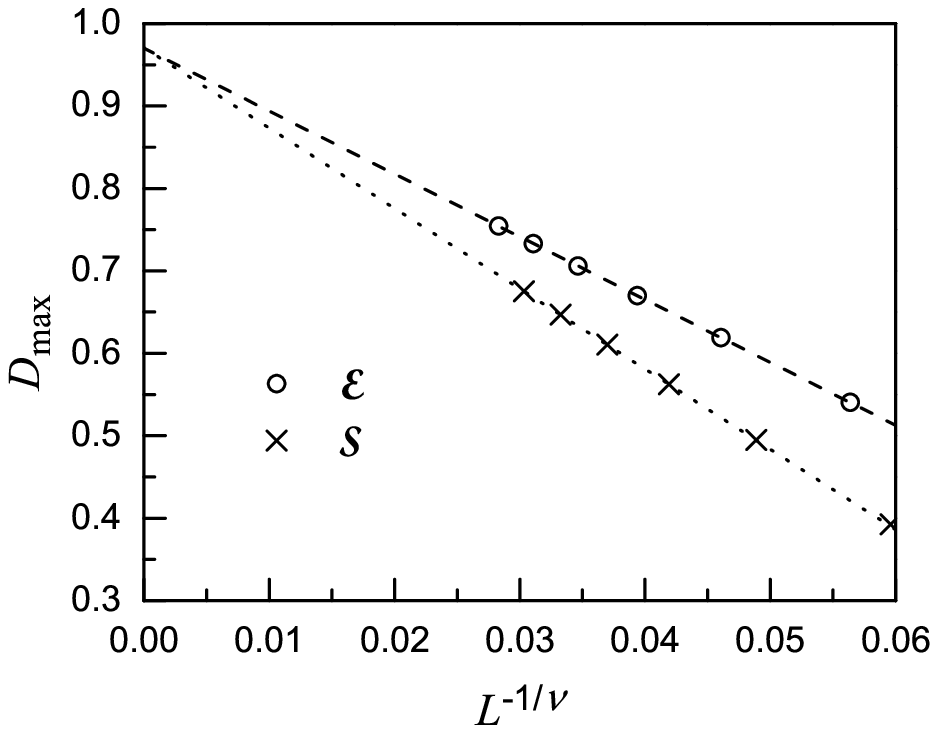}
\includegraphics[width=3in,height=2in]{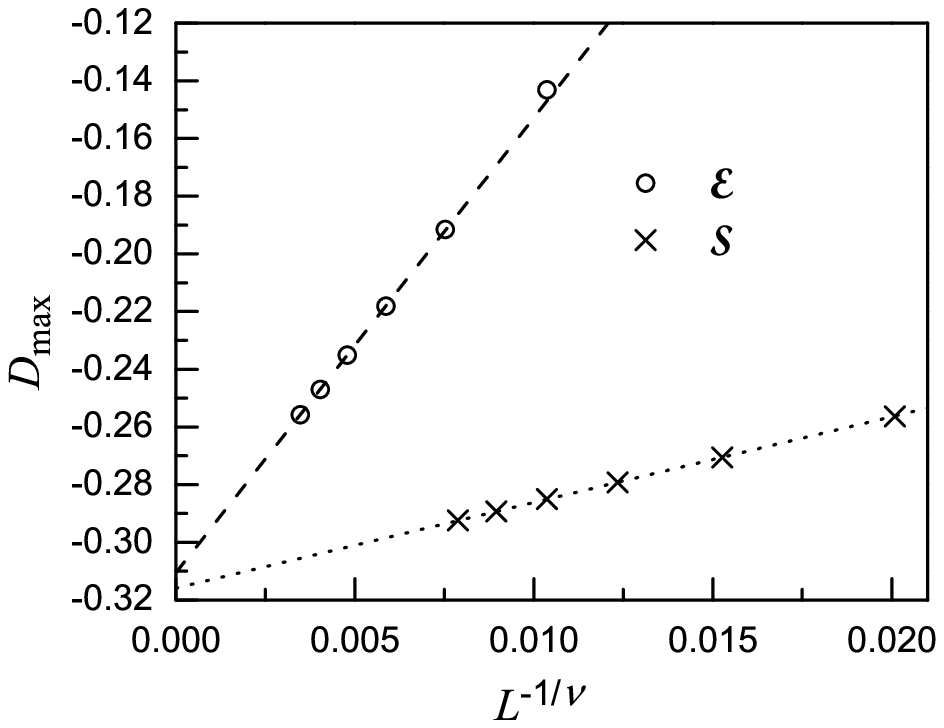}
\caption{Finite-size scaling of $D_{\rm max}$ versus $L^{-1/\nu}$.
The full lines are least square straight line fits for sizes with
$L \geq 100$. Top: the Haldane-Large-$D$ transition, where
$\nu\simeq 1.42$ ($\nu\simeq 1.45$) for those $D_{\rm max}$'s
corresponding to the local maxima in the curves of ${\cal E}$
(${\cal S}$). Bottom: the Haldane-N\'{e}el transition, where
$\nu\simeq 0.90$ ($\nu\simeq 1.05$) for those $D_{\rm max}$'s
corresponding to the local maxima in the curves of ${\cal E}$
(${\cal S}$). } \label{fig:crit_point}
\end{figure}
%%%%%%%%%%%%%%%%%%%%%%%%%%%%%%%%%%%%%%%%%%%%%%%%%
%%%%%%%%%%%%%%%%%%%%%%%%%%%%%%%%%%%%%%%%%%%%%%%%%
\begin{figure}
\includegraphics[width=3in]{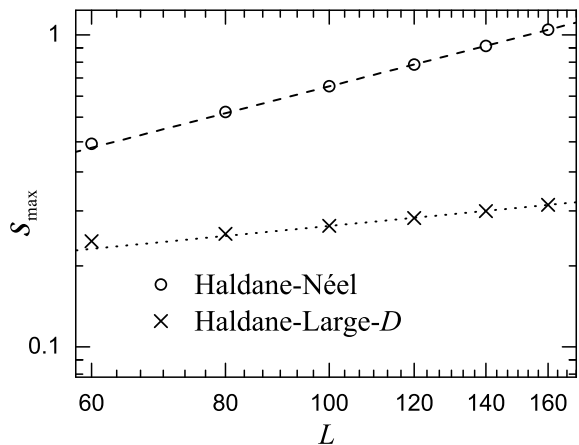}
\caption{The log-log plot of ${\cal S}_{\rm max}$ for various
sizes $L$. The full lines are least square straight line fits for
sizes with $L \geq 100$.} \label{fig:Smax}
\end{figure}
%%%%%%%%%%%%%%%%%%%%%%%%%%%%%%%%%%%%%%%%%%%%%%%%%

According to the finite-size scaling theory,~\cite{fss} one has
\begin{equation}
|D_{\rm max}(L) - D_c|  \propto L^{-1/\nu} \; , \label{eq:fss}
\end{equation}
where $D_c$ is the critical point in the thermodynamic limit and
$\nu$ is the critical exponent of the correlation length. Thus
$D_c$ can be determined by an extrapolation procedure. The results
for the Haldane-Large-$D$ and the Haldane-N\'{e}el transitions are
shown in Fig.~\ref{fig:crit_point}. For the Gaussian transition
between the Haldane and the large-$D$ phases, both extrapolations
give $D_c \simeq 0.97$ as shown in the top panel of
Fig.~\ref{fig:crit_point}. The critical exponent of the
correlation $\nu\simeq 1.42$ for the data of $D_{\rm max}(L)$
obtained from ${\cal E}$, while $\nu\simeq 1.45$ for those from
${\cal S}$. Because of the relation $\nu=1/(2-K)$, the Luttinger
liquid parameter $K=1.30$ ($K=1.31$) for the data related to
${\cal E}$ (${\cal S}$). We find that the values obtained from the
measurements of ${\cal E}$ and ${\cal S}$ agrees each other, and
they are consistent with the previous
findings,~\cite{DegliEspostiBoschi03,note4} where $D_c\simeq 0.99$
and $K\simeq 1.328$. For the Ising transition between the N\'{e}el
and the Haldane phases, $D_c\simeq -0.31$ for both extrapolations
as shown in the bottom panel of Fig.~\ref{fig:crit_point}. The
critical exponent of the correlation $\nu\simeq 0.90$ for the data
of $D_{\rm max}$ obtained from ${\cal E}$, while $\nu\simeq 1.05$
for those from ${\cal S}$. Again, the value of $D_c$ agrees with
the previous result ($D_c=-0.31$),~\cite{CamposVenuti06-2,note4}
and our findings of $\nu$ are consistent with the theoretical
prediction ($\nu=1$) for the Ising transition. From the above
discussions, we find that both the entanglement entropy and the
fidelity susceptibility are equally suited for revealing the
critical behaviors in the present case.

To verify the predicted critical scaling behavior of the fidelity
susceptibility in Eq.~(\ref{scaling_S}), the values ${\cal S}_{\rm
max}(L)$ of the local maxima for various sizes $L$ are plotted in
Fig.~\ref{fig:Smax}. It is found that our data do fulfill the
scaling relation in Eq.~(\ref{scaling_S}), where
$\Delta_{Q}=-0.33$ (i.e., $\Delta_V=1.34$) for the Gaussian
transition and $\Delta_{Q}=-0.89$ (i.e., $\Delta_V=1.06$) for the
Ising one ($d=1$ and $z=1$ are assumed here). The value of
$\Delta_V$ for the Ising transition agrees with the predicted one,
$\Delta_V=1$. Since $\Delta_V=K$ for the Gaussian transition, the
Luttinger liquid parameter $K$ determined by the present
finite-size scaling agrees with the previous
findings~\cite{DegliEspostiBoschi03} and those determined by the
critical exponent $\nu$ coming from the scaling in
Fig.~\ref{fig:crit_point}. Thus the fact that a single parameter
$K$ controls all the critical exponents for the Gaussian
transition is confirmed by our numerical results.

%%%%%%%%%%%%%%%%%%%%%%%%%%%%%%%%%%%%%%%%%%%%%%%%%%%%%%
%\section{Conclusions}
%%%%%%%%%%%%%%%%%%%%%%%%%%%%%%%%%%%%%%%%%%%%%%%%%%%%%%

In summary, the general scaling analysis of the fidelity
susceptibility proposed in Ref.~\onlinecite{CVZ07} is verified by
the present DMRG calculations for the model of Eq.~(\ref{hamilt}).
The critical points of the Gaussian and the Ising transitions, as
well as some of their critical exponents, are determined from the
perspective of quantum information. We note that, as seen from
Figs.~\ref{fig:crit_point} and \ref{fig:Smax}, data for systems of
smaller sizes can deviate from the fitting lines obtained from the
data for those of larger sizes (say, $L\geq 100$). Therefore, to
avoid the finite-size effects and to unveil the correct scaling
behaviors at the critical points, calculations for systems of
large enough sizes are necessary. From our DMRG calculation for
systems of large sizes, we conclude that the fidelity
susceptibility and the entanglement entropy can have similar
predictive power for revealing QPTs.

%%%%%%%%%%%%%%%%%%%%%%%%%%%%%%%%%%%%%%%%%%%%%%%%%%%%%%%%%%%%%%
%\begin{acknowledgments}
The authors are grateful to Hsiang-Hsuan Hung and Fabian
Heidrich-Meisner for many valuable discussions. This work was
supported by the National Science Council of Taiwan under Contract
No. NSC 96-2112-M-029-004-MY3.
%\end{acknowledgments}
%%%%%%%%%%%%%%%%%%%%%%%%%%%%%%%%%%%%%%%%%%%%%%%%%%%%%%%%%%%%%%


\begin{thebibliography}{99}
\bibitem{note1}
Here we use the notations of
Ref.~\onlinecite{DegliEspostiBoschi03}.
\bibitem{note2}
To lift the double degeneracy in the N\'{e}el phase, a Zeeman term
$h_1 S^z_1$ acting only upon the spin at site 1 is added in the
Hamiltonian of Eq.~(\ref{hamilt}). Here we take the boundary
magnetic field $h_1=-0.01$. The inclusion of this tiny Zeeman
splitting selects a specific ground state and stabilizes our
calculations of the fidelity in the N\'{e}el phase.
\bibitem{note4}
We note that the DMRG calculations in
Refs.~\onlinecite{DegliEspostiBoschi03} and
\onlinecite{CamposVenuti06-2} are under the periodic boundary
conditions, rather than the open boundary conditions used in the
present work. The small discrepancy of our results from theirs may
be due to the different boundary conditions employed.
\end{thebibliography}
\end{document}